\documentclass[12pt,executivepaper,openbib]{article}
\usepackage{amsmath}
\usepackage{amssymb}
\usepackage{amsfonts}

\input{tcilatex}

\begin{document}

\begin{center}
{\Large Weak-Pseudo-Hermiticity of Non-Hermitian Hamiltonians\ with
Position-Dependent Mass}

\bigskip

S.-A. Yahiaoui, M. Bentaiba\footnote{%
Corresponding author:
\par
E-mail address : bentaiba@hotmail.com
\par
\ \ \ \ \ \ \ \ \ \ \ \ \ \ \ \ \ \ \ \ \ \ bentaiba1@caramail.com}

LPTHIRM, D\'{e}partement de Physique, Facult\'{e} des Sciences,

Universit\'{e} Saad DAHLAB de Blida, Algeria.

\bigskip

\textbf{Abstract}
\end{center}

We extend the definition of $\eta -$weak-pseudo-Hermiticity to the class of
potentials endowed with position-dependent mass. The construction of
non-Hermitian Hamiltonians through some generating function are obtained.
Special cases of potentials are thus deduced.

\bigskip


$\ \ \ \ \ \ \ \ \ \ \ \ \ \ \ \mathcal{PT}-$symmetry; Effective mass.


\bigskip

\section{Introduction}

The Hamiltonians are called $\mathcal{PT}-$invariant if they are invariant
under a joint transformation of parity $\mathcal{P}$ and time-reversal $%
\mathcal{T}$ [1-8]. A conjecture due to Bender and Boettcher [1] has relaxed 
$\mathcal{PT}-$symmetry as a necessary condition for the reality of the
spectrum. Here, the Hermiticity assumption $\mathcal{H}=\mathcal{H}^{\dag }$
is replaced by the $\mathcal{PT}-$symmetric one; i.e. $\left[ \mathcal{PT},%
\mathcal{H}\right] =0$, where $\mathcal{P}$ denotes the parity operator
(space reflection) and has as effects : $x\rightarrow -x$, $p\rightarrow -p$
and $\mathcal{T}$ mimics the time-reversal and has as effects : $%
x\rightarrow x$, $p\rightarrow -p$, and $i\rightarrow -i$. Note that $%
\mathcal{T}$ changes the sign of $i$ because it preserves the fundamental
commutation relation of the quantum mechanics known as the Heisenberg
algebra, i.e. $\left[ x,p\right] =i\hbar $ [1-3].

According to Mostafazadeh [9-12], the basic mathematical structure
underlying the properties of $\mathcal{PT-}$symmetry is explored and can now
be found to be connected to the concept of a pseudo-Hermiticity. The
pseudo-Hermiticity has been found to be a more general concept then those of
Hermiticity and $\mathcal{PT}-$symmetry. As a consequence of this, the
reality of the bound-state eigenvalues can be associated with it.

In terms of these settings, a Hamiltonian $\mathcal{H}$ is called
pseudo-Hermitian if it obeys to [9,11]%
\begin{equation}
\mathcal{H}^{\dag }=\eta \mathcal{H}\eta ^{-1},  \tag{1}
\end{equation}%
where $\eta $ is a Hermitian invertible linear operator and a dagger $\left(
^{\dag }\right) $ stands for the adjoint of the corresponding operator. A
non-Hermitian Hamiltonian has a real spectrum if and only if it is
pseudo-Hermitian with respect to a linear Hermitian automorphism [10], and
may be factored as%
\begin{equation}
\eta =\mathcal{D}^{\dag }\mathcal{D},  \tag{2}
\end{equation}%
where $\mathcal{D}$ $:$ $\mathfrak{H\rightarrow H}$ is a linear automorphism
($\mathfrak{H}$ is the Hilbert space). Note that choosing $\eta =1$ reduces
the assumption (1) to the Hermiticity of the Hamiltonian.

On the other hand, Bagchi and Quesne [13] have established that the twin
concepts of pseudo-Hermiticity and weak-pseudo-Hermiticity are complementary
to one another. In the pseudo-Hermiticity case, $\eta $ can be written as a
first-order differential operator and may be anti-Hermitian, while in the
weak-pseudo-Hermitian case, $\eta $ is a second-order differential operator
and must be necessarily Hermitian.

The quantum mechanical systems with position-dependent mass have attracted,
in recent years, much attention on behalf of physicists [15-20]. The
effective mass Schr\"{o}dinger equation was first introduced by BenDaniel
and Duke in order to explain the behaviors of electrons in semi-conductors
[15]. It also have many applications in the fields of materials science and
condensed matter physics [20,21].

In the present paper, a class of non-Hermitian Hamiltonians, known in the
literature, as well as their accompanying ground-state wavefunctions are
generated as a by-product of the generalized $\eta -$weak-pseudo-Hermiticity
endowed with position-dependent mass. Here our primary concern is to point
out that, being different from the realization of Ref.[13] considering
therein $A\left( x\right) $ as a pure imaginary function, there is no
inconsistency if a shift on the momentum $p$ of the type $p\rightarrow p-%
\frac{A\left( x\right) }{U\left( x\right) }$ is used, where $A$ $\left(
x\right) $ and $U\left( x\right) \left( \neq 0\right) $ are, respectively,
complex- and real-valued functions. It opens a way towards the construction
of non-Hermitian Hamiltonians (not necessarily $\mathcal{PT}-$symmetric). On
these settings, Eq.(2) becomes $\eta \rightarrow \widetilde{\eta }=%
\widetilde{\mathcal{D}}^{\dag }\widetilde{\mathcal{D}}$. Such operator, i.e. 
$\widetilde{\mathcal{D}}$, may be looked upon as a gauge-transformed version
of $\mathcal{D}$, depending essentially on the function $A\left( x\right) $.
Consequently, it is found that the wavefunction is also subjected to a gauge
transformation of the type $\psi \left( x\right) \rightarrow \mathcal{\xi }%
\left( x\right) =\Lambda \left( x\right) \psi \left( x\right) $ where $%
\Lambda \left( x\right) =\exp \left[ i\int^{x}dy\frac{A\left( y\right) }{%
U\left( y\right) }\right] $.

\section{Generalized pseudo-Hermitian Hamiltonians}

The general form of the Hamiltonian introduced by von Roos [16] for the
spatially varying mass $M\left( x\right) =m_{0}m\left( x\right) $ reads%
\begin{equation}
\mathcal{H=}\frac{1}{4}\left[ m^{\alpha }\left( x\right) pm^{\beta }\left(
x\right) pm^{\gamma }\left( x\right) +m^{\gamma }\left( x\right) pm^{\beta
}\left( x\right) pm^{\alpha }\left( x\right) \right] +V\left( x\right) , 
\tag{3}
\end{equation}%
where the constraint $\alpha +\beta +\gamma =-1$ holds and $V\left( x\right)
=V_{\func{Re}}\left( x\right) +iV_{\func{Im}}\left( x\right) $ is a
complex-valued potential. Here, $p\left( =-i\frac{d}{dx}\right) $ is a
momentum with $\hbar =m_{0}=1$, and $m\left( x\right) $ is dimensionless
real-valued mass function.

Using the restricted Hamiltonian from the $\alpha =\gamma =0$ and $\beta =-1$
constraints, the Hamiltonian (3) becomes%
\begin{equation}
\mathcal{H}=pU^{2}\left( x\right) p+V\left( x\right) ,  \tag{4}
\end{equation}%
with $U^{2}\left( x\right) =\frac{1}{2m\left( x\right) }$. The shift on the
momentum $p$ in the manner%
\begin{equation}
p\rightarrow p-\frac{A\left( x\right) }{U\left( x\right) },  \tag{5}
\end{equation}%
where $A$ $:$ $%
\mathbb{R}
\rightarrow 
\mathbb{C}
$ is a complex-valued function, allows to bring the Hamiltonian of Eq.(4) in
the form%
\begin{equation}
\mathcal{H\rightarrow H}^{\prime }=\left[ p-\frac{A\left( x\right) }{U\left(
x\right) }\right] U^{2}\left( x\right) \left[ p-\frac{A\left( x\right) }{%
U\left( x\right) }\right] +V\left( x\right) .  \tag{6}
\end{equation}

In Ref.[11], it was showed that for every anti-pseudo-Hermitian Hamiltonian $%
\mathcal{H}$, there is an antilinear operator $\tau $ fulfilling the
condition%
\begin{equation}
\mathcal{H}^{\dag }=\tau \mathcal{H}\tau ^{-1}.  \tag{7}
\end{equation}

Let us extend the proof of Ref.[12] to our Hamiltonian (6). To this end, $%
\tau $ should be constructed suitably. According to Mostafazadeh [12], $\tau
=\mathcal{T}\func{e}^{i\alpha \left( x\right) }$ is the product of linear
and antilinear operators, and $\alpha $ $:$ $%
\mathbb{R}
\rightarrow 
\mathbb{C}
$ is a complex-valued function. Therefore, the Hermiticity of $\tau $ is
established straightforwardly

\begin{equation}
\tau ^{\dag }=\func{e}^{-i\alpha ^{\ast }\left( x\right) }\mathcal{T}^{\dag
}=\func{e}^{-i\alpha ^{\ast }\left( x\right) }\mathcal{T}=\mathcal{T}\func{e}%
^{i\alpha \left( x\right) }=\tau ,  \tag{8}
\end{equation}%
where the identities $\mathcal{T}^{\dag }\mathcal{=T}$ and $\mathcal{T}%
f\left( x\right) \mathcal{T}=f^{\ast }\left( x\right) $ are used and $f$ $:$ 
$%
\mathbb{R}
\rightarrow 
\mathbb{C}
$.

According to Mostafazadeh in Ref.[12], the function $\alpha \left( x\right) $
can be generalized to $\alpha \left( x\right) =-2\int^{x}dy\frac{A\left(
y\right) }{U\left( y\right) }$, therefore%
\begin{eqnarray}
\tau \mathcal{H}^{\prime }\tau ^{-1} &=&\mathcal{T}\func{e}^{i\alpha \left(
x\right) }\left[ p-\frac{A\left( x\right) }{U\left( x\right) }\right]
U^{2}\left( x\right) \left[ p-\frac{A\left( x\right) }{U\left( x\right) }%
\right] \func{e}^{-i\alpha \left( x\right) }\mathcal{T}  \notag \\
&&\mathcal{+T}\func{e}^{i\alpha \left( x\right) }V\left( x\right) \func{e}%
^{-i\alpha \left( x\right) }\mathcal{T}  \notag \\
&=&\mathcal{T}\left[ p-\frac{A\left( x\right) }{U\left( x\right) }-\partial
_{x}\alpha \right] \func{e}^{i\alpha \left( x\right) }U^{2}\left( x\right) 
\func{e}^{-i\alpha \left( x\right) }\left[ p-\frac{A\left( x\right) }{%
U\left( x\right) }-\partial _{x}\alpha \right] \mathcal{T}  \notag \\
&&\mathcal{+}V^{\ast }\left( x\right)  \notag \\
&=&\mathcal{T}\left[ p-\frac{A\left( x\right) }{U\left( x\right) }-\partial
_{x}\alpha \right] U^{2}\left( x\right) \left[ p-\frac{A\left( x\right) }{%
U\left( x\right) }-\partial _{x}\alpha \right] \mathcal{T+}V^{\ast }\left(
x\right)  \notag \\
&=&\mathcal{T}\left[ p+\frac{A\left( x\right) }{U\left( x\right) }\right]
U^{2}\left( x\right) \left[ p+\frac{A\left( x\right) }{U\left( x\right) }%
\right] \mathcal{T+}V^{\ast }\left( x\right)  \notag \\
&=&\left[ -p+\frac{A^{\ast }\left( x\right) }{U\left( x\right) }\right]
U^{2}\left( x\right) \left[ -p+\frac{A^{\ast }\left( x\right) }{U\left(
x\right) }\right] \mathcal{+}V^{\ast }\left( x\right)  \notag \\
&=&\left[ p-\frac{A^{\ast }\left( x\right) }{U\left( x\right) }\right]
U^{2}\left( x\right) \left[ p-\frac{A^{\ast }\left( x\right) }{U\left(
x\right) }\right] \mathcal{+}V^{\ast }\left( x\right)  \notag \\
&=&\mathcal{H}^{\prime \dag },  \TCItag{9}
\end{eqnarray}%
where for every differential function $\alpha \left( x\right) $, the
following identity holds $\func{e}^{-i\alpha \left( x\right) }p\func{e}%
^{i\alpha \left( x\right) }=p+\partial _{x}{}\alpha \left( x\right) $ while
the position $x$ commutes with $\func{e}^{i\alpha \left( x\right) }$ and
remains unaffected under a last transformation; i.e. $\func{e}^{-i\alpha
\left( x\right) }x\func{e}^{i\alpha \left( x\right) }=x.$ Here we note that
for every function $f$ $:$ $%
\mathbb{R}
\rightarrow 
\mathbb{C}
$, the identity $\mathcal{T}f\left( x,p\right) \mathcal{T}=f^{\ast }\left(
x,-p\right) $ is used.

In the other hand, and according to Ref.[11], it was checked that $\mathcal{%
PT}-$symmetry $\left( \left[ \mathcal{PT},\mathcal{H}\right] =0\right) $ and
anti-pseudo-Hermiticity operator $\tau $ imply pseudo-Hermiticity of $%
\mathcal{H}$ with the respect of a linear Hermitian automorphism $\eta $ $:$ 
$\mathfrak{H\rightarrow H}$ according to%
\begin{equation}
\eta =\tau \mathcal{PT},  \tag{10}
\end{equation}%
and it turns out that the choice of $\eta $ is not unique. As was made for $%
\tau $, let us generalize $\eta $ according to%
\begin{equation}
\eta =\exp \left[ 2i\int^{x}dy\frac{A^{\ast }\left( y\right) }{U\left(
y\right) }\right] \mathcal{P},  \tag{11}
\end{equation}%
then the Hermiticity of $\eta $ is established straightforwardly%
\begin{eqnarray}
\eta ^{\dag } &=&\mathcal{P}\exp \left[ -2i\int^{x}dy\frac{A\left( y\right) 
}{U\left( y\right) }\right] =\exp \left[ -2i\int^{-x}dy\frac{A\left(
y\right) }{U\left( y\right) }\right] \mathcal{P}  \notag \\
&=&\exp \left[ 2i\int^{-x}d\left( -y\right) \frac{A\left( y\right) }{U\left(
y\right) }\right] \mathcal{P=}\exp \left[ 2i\int^{x}dy\frac{A\left(
-y\right) }{U\left( -y\right) }\right] \mathcal{P}  \notag \\
&=&\exp \left[ 2\int^{x}dy\frac{i\func{Re}A\left( -y\right) -\func{Im}%
A\left( -y\right) }{U\left( -y\right) }\right] \mathcal{P}  \notag \\
&=&\exp \left[ 2\int^{x}dy\frac{i\func{Re}A\left( y\right) +\func{Im}A\left(
y\right) }{U\left( y\right) }\right] \mathcal{P}  \notag \\
&=&\exp \left[ 2i\int^{x}dy\frac{\func{Re}A\left( y\right) -i\func{Im}%
A\left( y\right) }{U\left( y\right) }\right] \mathcal{P}  \notag \\
&=&\exp \left[ 2i\int^{x}dy\frac{A^{\ast }\left( y\right) }{U\left( y\right) 
}\right] \mathcal{P}  \notag \\
&=&\eta ,  \TCItag{12}
\end{eqnarray}%
where we use $\mathcal{P}^{\dag }=\mathcal{P}$ and, for every function $f$ $%
: $ $%
\mathbb{R}
\rightarrow 
\mathbb{C}
$, the following identity holds $\mathcal{P}f\left( x\right) \mathcal{P}%
=f\left( -x\right) $. In Eq.(12), the real and imaginary parts of $A\left(
x\right) $ are, respectively, even and odd functions; i.e. $\func{Re}A\left(
-x\right) =\func{Re}A\left( x\right) $, $\func{Im}A\left( -x\right) =-\func{%
Im}A\left( x\right) $ and $U\left( x\right) $ must be an even function, i.e. 
$U\left( x\right) =U\left( -x\right) $.

In summary, the $\mathcal{PT}-$symmetry and anti-pseudo-Hermiticity with
respect to $\tau $ imply pseudo-Hermiticity with respect to $\tau \mathcal{PT%
}$ and which coincides with the $\eta $ operator [11]. Therefore, it is
obvious that the (weak-) pseudo-Hermiticity as defined in Eq.(10) adapts
very well to the problems relating with position-dependent effective mass.

\section{The generalized weak-pseudo-Hermiticity generators}

As $\eta $ is weak-pseudo-Hermitian, then the operators $\mathcal{D}$ and $%
\mathcal{D}^{\dag }$ are connected to the first-order differential operator
through [14]%
\begin{eqnarray}
\mathcal{D} &=&U\left( x\right) \partial _{x}+\phi \left( x\right) ,  \notag
\\
&=&iU\left( x\right) p+\phi \left( x\right) ,  \TCItag{13.a} \\
\mathcal{D}^{\dag } &=&-\partial _{x}U\left( x\right) +\phi ^{\ast }\left(
x\right) ,  \notag \\
&=&-ipU\left( x\right) +\phi ^{\ast }\left( x\right) ,  \TCItag{13.b}
\end{eqnarray}%
where we have used the abbreviation $\partial _{x}=\frac{d}{dx}$. Here $\phi 
$ $:$ $%
\mathbb{R}
\rightarrow 
\mathbb{C}
$ is a complex-valued function. It is obvious that the operator $\mathcal{D}$
becomes, under transformation (5),%
\begin{eqnarray}
\widetilde{\mathcal{D}} &=&iU\left( x\right) \left[ p-\frac{A\left( x\right) 
}{U\left( x\right) }\right] +\phi \left( x\right) ,  \notag \\
&=&iU\left( x\right) p-iA\left( x\right) +\phi \left( x\right) .  \TCItag{14}
\end{eqnarray}

Therefore, the operator $\widetilde{\mathcal{D}}$ may be looked upon as a
gauge-transformed version of $\mathcal{D}$, depending on $A\left( x\right) $
such that $\widetilde{D}=\mathcal{D-}iA\left( x\right) $. In terms of these, 
$\widetilde{\eta }$ becomes%
\begin{eqnarray}
\widetilde{\eta } &=&\widetilde{\mathcal{D}}^{\dag }\widetilde{\mathcal{D}} 
\notag \\
&=&\left[ \mathcal{D}^{\dag }+iA^{\ast }\left( x\right) \right] \left[ 
\mathcal{D}-iA\left( x\right) \right]  \notag \\
&=&\mathcal{D}^{\dag }\mathcal{D}-i\mathcal{D}^{\dag }A\left( x\right)
+iA^{\ast }\left( x\right) \mathcal{D}+A^{\ast }\left( x\right) A\left(
x\right) ,  \TCItag{15}
\end{eqnarray}%
and taking into account that $\phi \left( x\right) =f\left( x\right)
+ig\left( x\right) $ and $A\left( x\right) =a\left( x\right) +ib\left(
x\right) $, (15) can be recast as%
\begin{eqnarray}
\widetilde{\eta } &=&\mathcal{D}^{\dag }\mathcal{D}+2iU\left( x\right)
a\left( x\right) \partial _{x}+i\left[ U\left( x\right) A\left( x\right) %
\right] ^{\prime }-i\phi ^{\ast }\left( x\right) A\left( x\right)  \notag \\
&&+i\phi \left( x\right) A^{\ast }\left( x\right) +\left\vert A\left(
x\right) \right\vert ^{2},  \TCItag{16}
\end{eqnarray}%
where prime denotes derivative with respect to $x$. At this point, let us
now evaluate $\eta $ appearing in Eq.(16) using Eq.(13), we obtain%
\begin{eqnarray}
\mathcal{D}^{\dag }\mathcal{D} &=&\left[ -\partial _{x}U\left( x\right)
+\phi \left( x\right) \right] \left[ U\left( x\right) \partial _{x}+\phi
\left( x\right) \right]  \notag \\
&=&-U^{2}\left( x\right) \partial _{x}^{2}-2U\left( x\right) \left[
U^{\prime }\left( x\right) +ig\left( x\right) \right] \partial
_{x}+\left\vert \phi \left( x\right) \right\vert ^{2}  \notag \\
&&-\left[ U\left( x\right) \phi \left( x\right) \right] ^{\prime }, 
\TCItag{17}
\end{eqnarray}

Combining Eq.(17) with Eq.(16), we obtain a second-order differential
operator of $\widetilde{\eta }$%
\begin{equation}
\widetilde{\eta }=-U^{2}\left( x\right) \partial _{x}^{2}-2\mathcal{K}\left(
x\right) \partial _{x}+\mathcal{L}\left( x\right) ,  \tag{18}
\end{equation}%
where $\mathcal{K}\left( x\right) $ and $\mathcal{L}\left( x\right) $ are
defined as%
\begin{eqnarray}
\mathcal{K}\left( x\right)  &=&U\left( x\right) U^{\prime }\left( x\right)
+iU\left( x\right) g\left( x\right) -iU\left( x\right) a\left( x\right) , 
\TCItag{19.a} \\
\mathcal{L}\left( x\right)  &=&\left\vert \phi \left( x\right) \right\vert
^{2}+\left\vert A\left( x\right) \right\vert ^{2}-\left[ U\left( x\right)
\phi \left( x\right) \right] ^{\prime }+i\left[ U\left( x\right) A\left(
x\right) \right] ^{\prime }  \notag \\
&&-i\phi ^{\ast }\left( x\right) A\left( x\right) +i\phi \left( x\right)
A^{\ast }\left( x\right) .  \TCItag{19.b}
\end{eqnarray}

One can easily check that $\widetilde{\eta }$ given in Eq.(18) is, indeed,
Hermitian since it is written in the form $\widetilde{\eta }=\widetilde{%
\mathcal{D}}^{\dag }\widetilde{\mathcal{D}}$. On the other hand, taking into
account $p=-i\partial _{x}$, the Hamiltonian of Eq.(6) may be expressed as%
\begin{equation}
\mathcal{H}^{\prime }=-U^{2}\left( x\right) \partial _{x}^{2}-2\mathcal{M}%
_{1}\left( x\right) \partial _{x}+\mathcal{N}_{1}\left( x\right) +V\left(
x\right) ,  \tag{20}
\end{equation}%
where, by definition%
\begin{eqnarray}
\mathcal{M}_{1}\left( x\right)  &=&U\left( x\right) U^{\prime }\left(
x\right) -iU\left( x\right) A\left( x\right) ,  \TCItag{21.a} \\
\mathcal{N}_{1}\left( x\right)  &=&i\left[ U\left( x\right) A\left( x\right) %
\right] ^{\prime }+A^{2}\left( x\right) .  \TCItag{21.b}
\end{eqnarray}

The adjoint of the Hamiltonian (20) reads as%
\begin{equation}
\mathcal{H}^{\prime \dag }=-U^{2}\left( x\right) \partial _{x}^{2}-2\mathcal{%
M}_{2}\left( x\right) \partial _{x}+\mathcal{N}_{2}\left( x\right) +V^{\ast
}\left( x\right) ,  \tag{22}
\end{equation}%
with%
\begin{eqnarray}
\mathcal{M}_{2}\left( x\right) &=&U\left( x\right) U^{\prime }\left(
x\right) -iU\left( x\right) A^{\ast }\left( x\right) ,  \TCItag{23.a} \\
\mathcal{N}_{2}\left( x\right) &=&i\left[ U\left( x\right) A^{\ast }\left(
x\right) \right] ^{\prime }+A^{\ast 2}\left( x\right) .  \TCItag{23.b}
\end{eqnarray}

It should be noted that $\mathcal{D}$ and $\mathcal{D}^{\dag }$ are two
intertwining operators, therefore, the defining condition (1) may be
expressed as $\eta \mathcal{H}=\mathcal{H}^{\dag }\eta $. Thereupon, a
generalization beyond the pair $\widetilde{\eta }$ and $\mathcal{H}^{\prime
} $ is straightforward, given%
\begin{equation}
\widetilde{\eta }\mathcal{H}^{\prime }=\mathcal{H}^{\prime \dag }\widetilde{%
\eta }.  \tag{24}
\end{equation}

Letting both sides of (24) act on every function, e.g. \ on a wavefunction.
Using Eqs.(18), (20), (22) and comparing between their varying differential
coefficients, we can easily recognized from the coefficients corresponding
to the third derivative that $A\left( x\right) $ must be real function, i.e. 
$b\left( x\right) =0$.

By comparing both coefficients corresponding to the second derivative, one
deduces the expression connecting the potential to its conjugate through%
\begin{equation}
V\left( x\right) =V^{\ast }\left( x\right) -4iU\left( x\right) g^{\prime
}\left( x\right) .  \tag{25}
\end{equation}

On the other hand, the coefficients corresponding to the first derivative
give the shape of the potential%
\begin{equation}
V^{\ast \prime }\left( x\right) =2f\left( x\right) f^{\prime }\left(
x\right) -2g\left( x\right) g^{\prime }\left( x\right) -\left[ U\left(
x\right) f\left( x\right) \right] ^{\prime \prime }+2i\left[ U\left(
x\right) g^{\prime }\left( x\right) \right] ^{\prime },  \tag{26}
\end{equation}%
and by integrating Eq.(26) taking into account its conjugate, we get%
\begin{eqnarray}
V\left( x\right)  &\equiv &V_{\func{Re}}\left( x\right) +iV_{\func{Im}%
}\left( x\right)   \notag \\
&=&f^{2}\left( x\right) -g^{2}\left( x\right) -\left[ U\left( x\right)
f\left( x\right) \right] ^{\prime }-2iU\left( x\right) g^{\prime }\left(
x\right) +\delta ,  \TCItag{27}
\end{eqnarray}%
with $\delta $ is a constant of integration. It is obvious that both
imaginary parts of Eqs.(25) and (27) coincide.

The last remaining coefficients correspond to the null derivative and give
the following pure-imaginary expression%
\begin{equation*}
-4U\left( x\right) f\left( x\right) f^{\prime }\left( x\right) g^{\prime
}\left( x\right) -4U\left( x\right) f^{2}\left( x\right) g^{\prime }\left(
x\right) +4U^{2}\left( x\right) f^{\prime }\left( x\right) g^{\prime }\left(
x\right)
\end{equation*}%
\begin{gather}
+4U\left( x\right) U^{\prime }\left( x\right) f^{\prime }\left( x\right)
g\left( x\right) +4U\left( x\right) U^{\prime }\left( x\right) f\left(
x\right) g^{\prime }\left( x\right) +2U^{2}\left( x\right) f^{\prime \prime
}\left( x\right) g\left( x\right)  \notag \\
+3U^{2}\left( x\right) U^{\prime }\left( x\right) g^{\prime \prime }\left(
x\right) +2U\left( x\right) U^{\prime \prime }\left( x\right) f\left(
x\right) g\left( x\right) -U^{2}\left( x\right) U^{\prime \prime }\left(
x\right) g^{\prime }\left( x\right)  \notag \\
-2U\left( x\right) U^{\prime }\left( x\right) U^{\prime \prime }\left(
x\right) g\left( x\right) +U^{3}\left( x\right) g^{\prime \prime \prime
}\left( x\right) -U^{2}\left( x\right) U^{\prime \prime \prime }\left(
x\right) g\left( x\right) =0.  \tag{28}
\end{gather}

Using Eq.(24) together with the eigenvalues of the Schr\"{o}dinger equation
for the Hamiltonian and its adjoint, namely $\mathcal{H}^{\prime }\left\vert
\xi _{i}\right\rangle =\mathcal{E}_{i}^{\prime }\left\vert \xi
_{i}\right\rangle $ and $\left\langle \xi _{j}\right\vert \mathcal{H}%
^{\prime \dag }=\left\langle \xi _{j}\right\vert \mathcal{E}_{j}^{\prime
\ast }$, where $\left\vert \xi _{q}\right\rangle \in \mathfrak{H}$ $\left(
q=i,j\right) $, and then multiplying them by $\widetilde{\eta }$ on the
left- and right-hand sides, respectively, we can easily obtain due to
Eq.(24), on subtracting, that any two eigenvectors $\left\vert \xi
_{i}\right\rangle $ and $\left\vert \xi _{j}\right\rangle $ satisfy%
\begin{eqnarray}
\left\langle \xi _{j}\right\vert \left( \mathcal{H}^{\prime \dag }\widetilde{%
\eta }-\widetilde{\eta }\mathcal{H}^{\prime }\right) \left\vert \xi
_{i}\right\rangle  &=&\left\langle \xi _{j}\right\vert \left( \mathcal{E}%
_{j}^{\prime \ast }\widetilde{\eta }-\mathcal{E}_{i}^{\prime }\widetilde{%
\eta }\right) \left\vert \xi _{i}\right\rangle   \notag \\
&=&\left( \mathcal{E}_{j}^{\prime \ast }-\mathcal{E}_{i}^{\prime }\right)
\left\langle \xi _{j}\right\vert \widetilde{\eta }\left\vert \xi
_{i}\right\rangle   \notag \\
&=&\left( \mathcal{E}_{j}^{\prime \ast }-\mathcal{E}_{i}^{\prime }\right)
\left\langle \xi _{j}\parallel \xi _{i}\right\rangle _{\widetilde{\eta }} 
\notag \\
&\equiv &0,  \TCItag{29}
\end{eqnarray}%
where $\left\langle \xi _{j}\parallel \xi _{i}\right\rangle _{\widetilde{%
\eta }}\equiv \left\langle \xi _{j}\right\vert \widetilde{\eta }\left\vert
\xi _{i}\right\rangle $ is the Hermitian indefinite inner product of the
Hilbert space $\mathfrak{H}$ defined by $\widetilde{\eta }$ [9,11].
According to the proposition 2 in Ref.[9], a direct implication of Eq.(29)
has the following properties

\begin{quotation}
(i) The eigenvectors with non-real eigenvalues have a vanishing $\eta -$%
norm, i.e. $\mathcal{E}_{i}^{\prime }\notin 
\mathbb{R}
$ implies that $\left\Vert \left\vert \xi _{i}\right\rangle \right\Vert _{%
\widetilde{\eta }}^{2}=\left\langle \xi _{i}\parallel \xi _{i}\right\rangle
_{\widetilde{\eta }}=0$.

(ii) Any two eigenvectors are $\eta -$orthogonal \textit{unless }their
eigenvalues are complex conjugates, i.e. $\mathcal{E}_{i}^{\prime }\neq 
\mathcal{E}_{j}^{\prime \ast }$ implies that $\left\langle \xi _{i}\parallel
\xi _{j}\right\rangle _{\widetilde{\eta }}=0$.
\end{quotation}

The inner product $\left\langle \text{\textperiodcentered }\parallel \text{%
\textperiodcentered }\right\rangle _{\widetilde{\eta }}$ is generally
positive-definite, i.e. $\left\langle \text{\textperiodcentered }\parallel 
\text{\textperiodcentered }\right\rangle _{\widetilde{\eta }}>0$. Thus, the
Hilbert space equipped with this inner product may be identified as the
physical Hilbert space $\mathfrak{H}_{\text{phys}}$ [1-3]. Therefore,
according to Eq.(29), it is obvious that $\mathcal{E}^{\prime }\mathcal{=E}%
^{\prime \ast }$. Hence, the eigenvalue $\mathcal{E}^{\prime }$ is real,
i.e. $\mathcal{E}_{\func{Im}}^{\prime }=0$. In terms of these, $\eta -$%
orthogonality suggests that the eigenvector (wavefunction), here $\mathcal{%
\xi }\left( x\right) $, is related to $\mathcal{H}^{\prime }$ through the
identity $\widetilde{\eta }\mathcal{\xi }\left( x\right) =0$ [14], i.e. 
\begin{equation}
\widetilde{\mathcal{D}}\mathcal{\xi }\left( x\right) =0,  \tag{30}
\end{equation}%
and keeping in mind Eq.(14), and after integration, we obtain the
ground-state wavefunction (not necessarily normalizable)%
\begin{eqnarray}
\mathcal{\xi }\left( x\right) &=&\Lambda \left( x\right) \psi \left( x\right)
\notag \\
&=&\exp \left[ i\int^{x}dy\frac{A\left( y\right) }{U\left( y\right) }\right]
\psi \left( x\right)  \notag \\
&\propto &\exp \left[ -\int^{x}dy\frac{f\left( y\right) }{U\left( y\right) }%
-i\int^{x}dy\frac{g\left( y\right) -a\left( y\right) }{U\left( y\right) }%
\right] ,  \TCItag{31}
\end{eqnarray}%
where $\psi \left( x\right) $ is the ground-state wavefunction when the
restriction $A\left( x\right) =0$ holds. Then $\xi \left( x\right) $, as for 
$\widetilde{\mathcal{D}}$, is also subjected to a gauge transformation in
the manner of $\psi \left( x\right) \rightarrow \mathcal{\xi }\left(
x\right) =\Lambda \left( x\right) \psi \left( x\right) $.

In these settings, letting $\widetilde{\mathcal{D}}$ acts on both sides of
(31), we obtain%
\begin{eqnarray}
\widetilde{\mathcal{D}}\xi \left( x\right) &\equiv &\left[ U\left( x\right)
\partial _{x}-iA\left( x\right) +\phi \left( x\right) \right] \Lambda \left(
x\right) \psi \left( x\right)  \notag \\
&=&U\left( x\right) \Lambda ^{\prime }\left( x\right) \psi \left( x\right)
+U\left( x\right) \Lambda \left( x\right) \psi ^{\prime }\left( x\right)
-iA\left( x\right) \Lambda \left( x\right) \psi \left( x\right)  \notag \\
&&+\phi \left( x\right) \Lambda \left( x\right) \psi \left( x\right)  \notag
\\
&=&\Lambda \left( x\right) \left[ U\left( x\right) \partial _{x}+\phi \left(
x\right) \right] \psi \left( x\right)  \notag \\
&\Longrightarrow &\mathcal{D\psi }\left( x\right) =0,  \TCItag{32}
\end{eqnarray}%
where $\Lambda ^{\prime }\left( x\right) =i\frac{A\left( x\right) }{U\left(
x\right) }\Lambda \left( x\right) $. That means that the wavefunctions thus
obtained can be deduced either by $\widetilde{\mathcal{D}}\xi \left(
x\right) =0$ or by $\mathcal{D}\psi \left( x\right) =0$.

In the remainder of the article, we write $\mathcal{E}$ instead of $\mathcal{%
E}^{\prime }$. Now, using the Schr\"{o}dinger equation $\mathcal{H}^{\prime }%
\mathcal{\xi }\left( x\right) =\mathcal{E\xi }\left( x\right) $, with $%
\mathcal{H}^{\prime }$ given in Eq.(20), $\mathcal{\xi }\left( x\right) $ in
Eq.(31) and $\mathcal{E}=\mathcal{E}_{\func{Re}}+i\mathcal{E}_{\func{Im}}$,
we end up by relating $f\left( x\right) $ to $g\left( x\right) $ and $%
U\left( x\right) $ through%
\begin{equation}
f\left( x\right) =\frac{U^{\prime }\left( x\right) g\left( x\right) -U\left(
x\right) g^{\prime }\left( x\right) }{2g\left( x\right) },  \tag{33}
\end{equation}%
where for the sake of simplicity we considere $\delta \equiv \mathcal{E}_{%
\func{Re}}$. Hence, it becomes clear that $g\left( x\right) $ is our
generating function leading to identify the function $f\left( x\right) $,
and then the potential $V\left( x\right) $.

This in turn leads to the following question. Is (33) the equation
connecting $f\left( x\right) $ to the generating function $g\left( x\right) $%
? The answer to this question amounts to check for the satisfaction of
Eq.(28). It is then straightforward, after a long calculation, to be
convinced that $f\left( x\right) $, as defined in (33), is a farfetched
function (solution).

In order to deal with position-dependent mass, we introduce the auxiliary
function defined by the mapping $\mu \left( x\right) \equiv \int^{x}\frac{dy%
}{U\left( y\right) }$, where $\mu \left( x\right) $ is a dimensionless mass
integral which will appear frequently in subsequent developments. The
function $f\left( x\right) $ can be written as%
\begin{equation}
f\left( x\right) =-\frac{g^{\prime }\left( x\right) }{2\mu ^{\prime }\left(
x\right) g\left( x\right) }-\frac{\mu ^{\prime \prime }\left( x\right) }{%
2\mu ^{\prime 2}\left( x\right) }\text{.}  \tag{34}
\end{equation}%
and the potential $V\left( x\right) $ acquires the form%
\begin{eqnarray}
V_{\text{eff}}\left( x\right) -\mathcal{E}_{\func{Re}} &=&-g^{2}\left(
x\right) -\frac{g^{\prime 2}\left( x\right) }{4g^{2}\left( x\right) \mu
^{\prime 2}\left( x\right) }+\frac{g^{\prime \prime }\left( x\right) }{%
2g\left( x\right) \mu ^{\prime 2}\left( x\right) }-\frac{g^{\prime }\left(
x\right) \mu ^{\prime \prime }\left( x\right) }{2g\left( x\right) \mu
^{\prime 3}\left( x\right) }  \notag \\
&&-2i\frac{g^{\prime }\left( x\right) }{\mu ^{\prime }\left( x\right) }, 
\TCItag{35}
\end{eqnarray}%
where $V_{\text{eff}}\left( x\right) $ is called the effective potential and
is related to $V\left( x\right) $ by%
\begin{equation}
V\left( x\right) =V_{\text{eff}}\left( x\right) -\mathcal{V}_{\mu }\left(
x\right) ,  \tag{36}
\end{equation}%
with%
\begin{equation}
\mathcal{V}_{\mu }\left( x\right) =\frac{\mu ^{\prime \prime \prime }\left(
x\right) }{\mu ^{\prime 3}\left( x\right) }-\frac{5}{4}\frac{\mu ^{\prime
\prime 2}\left( x\right) }{\mu ^{\prime 4}\left( x\right) }.  \tag{37}
\end{equation}

\section{Effective potentials and corresponding wavefunctions}

The strategy to determine both effective potentials and ground-state
wavefunctions is as follows. As $g\left( x\right) $ is a generating
function, all expressions depend on it. We may choose various generating
functions $g\left( x\right) $ and obtain all others expressions such as $%
f\left( x\right) $, $V_{\text{eff}}\left( x\right) $ and $\widetilde{\eta }$%
. Knowing $f\left( x\right) $ and $g\left( x\right) $, the proper
ground-state wavefunctions can be found from Eq.(32), i.e. without the
gauge-term. Without giving the details of our calculation which are
straightforward, we present the results of various expressions in standard
form.

\subsection{$3D-$Harmonic oscillator potential}

\begin{eqnarray}
g\left( x\right) &=&\alpha \mu \left( x\right) ,  \TCItag{38.a} \\
f\left( x\right) &=&-\frac{1}{2\mu \left( x\right) }-\frac{\mu ^{\prime
\prime }\left( x\right) }{2\mu ^{\prime 2}\left( x\right) },  \TCItag{38.b}
\\
V_{\text{HO}}\left( x\right) &=&-\alpha ^{2}\mu ^{2}\left( x\right) -\frac{1%
}{4\mu ^{2}\left( x\right) }-2i\alpha ,  \TCItag{38.c} \\
\psi _{\text{HO}}^{\left( 0\right) }\left( x\right) &\propto &\frac{\sqrt{%
\mu \left( x\right) }}{U\left( x\right) }\exp \left[ -\frac{i\alpha }{2}\mu
^{2}\left( x\right) \right] .  \TCItag{38.d}
\end{eqnarray}

\subsection{Morse potential}

\begin{eqnarray}
g\left( x\right) &=&\exp \left[ -\alpha \mu \left( x\right) \right] , 
\TCItag{39.a} \\
f\left( x\right) &=&\frac{\alpha }{2}-\frac{\mu ^{\prime \prime }\left(
x\right) }{2\mu ^{\prime 2}\left( x\right) },  \TCItag{39.b} \\
V_{\text{M}}\left( x\right) &=&-\exp \left[ -2\alpha \mu \left( x\right) %
\right] +2i\alpha \exp \left[ -\alpha \mu \left( x\right) \right] +\frac{%
\alpha ^{2}}{4},  \TCItag{39.c} \\
\psi _{\text{M}}^{\left( 0\right) }\left( x\right) &\propto &\frac{1}{%
U\left( x\right) }\func{e}^{-\frac{\alpha }{2}\mu \left( x\right) }\exp %
\left[ \frac{2i}{\alpha }\func{e}^{-\alpha \mu \left( x\right) }\left(
x\right) \right]  \TCItag{39.d}
\end{eqnarray}

\subsection{Scarf II potential}

\begin{eqnarray}
g\left( x\right) &=&\func{sech}\left[ \alpha \mu \left( x\right) \right] , 
\TCItag{40.a} \\
f\left( x\right) &=&\frac{\alpha }{2}\tanh \left[ \alpha \mu \left( x\right) %
\right] -\frac{\mu ^{\prime \prime }\left( x\right) }{2\mu ^{\prime 2}\left(
x\right) },  \TCItag{40.b} \\
V_{\text{Sc}}\left( x\right) &=&-\left( 1+\frac{3\alpha ^{2}}{4}\right) 
\func{sech}^{2}\left[ \alpha \mu \left( x\right) \right]  \notag \\
&&+2i\alpha \func{sech}\left[ \alpha \mu \left( x\right) \right] \tanh \left[
\alpha \mu \left( x\right) \right] +\frac{\alpha ^{2}}{4},  \TCItag{40.c} \\
\psi _{\text{Sc}}^{\left( 0\right) }\left( x\right) &\propto &\frac{1}{%
U\left( x\right) \sqrt{\cosh \left[ \alpha \mu \left( x\right) \right] }}%
\exp \left[ -\frac{i}{\alpha }\arctan \tanh \frac{\alpha }{2}\mu \left(
x\right) \right] .  \TCItag{40.d}
\end{eqnarray}

\subsection{Generalized P\"{o}schl-Teller potential}

\begin{eqnarray}
g\left( x\right) &=&\func{cosech}\left[ \alpha \mu \left( x\right) \right] ,
\TCItag{41.a} \\
f\left( x\right) &=&\frac{\alpha }{2}\coth \left[ \alpha \mu \left( x\right) %
\right] -\frac{\mu ^{\prime \prime }\left( x\right) }{\mu ^{\prime 2}\left(
x\right) },  \TCItag{41.b} \\
V_{\text{GPT}}\left( x\right) &=&-\left( 1-\frac{3\alpha ^{2}}{4}\right) 
\func{cosech}^{2}\left[ \alpha \mu \left( x\right) \right]  \notag \\
&&+2i\alpha \func{cosech}\left[ \alpha \mu \left( x\right) \right] \coth %
\left[ \alpha \mu \left( x\right) \right] +\frac{\alpha ^{2}}{4}, 
\TCItag{41.c} \\
\psi _{\text{GPT}}^{\left( 0\right) }\left( x\right) &\propto &\frac{1}{%
U\left( x\right) \sqrt{\sinh \left[ \alpha \mu \left( x\right) \right] }}%
\tanh ^{-\frac{2i}{\alpha }}\left[ \frac{\alpha \mu \left( x\right) }{2}%
\right] .  \TCItag{41.d}
\end{eqnarray}

\subsection{P\"{o}schl-Teller potential}

\begin{eqnarray}
g\left( x\right) &=&\func{sech}\left[ \alpha \mu \left( x\right) \right] 
\func{cosech}\left[ \alpha \mu \left( x\right) \right] ,  \TCItag{42.a} \\
f\left( x\right) &=&\alpha \coth \left[ 2\alpha \mu \left( x\right) \right] -%
\frac{\mu ^{\prime \prime }\left( x\right) }{2\mu ^{\prime 2}\left( x\right) 
},  \TCItag{42.b} \\
V_{\text{PT}}\left( x\right) &=&\left( \frac{3\alpha ^{2}}{4}-1+2i\alpha
\right) \func{cosech}^{2}\left[ \alpha \mu \left( x\right) \right]  \notag \\
&&-\left( \frac{3\alpha ^{2}}{4}-1-2i\alpha \right) \func{sech}^{2}\left[
\alpha \mu \left( x\right) \right] +\alpha ^{2},  \TCItag{42.c} \\
\psi _{\text{PT}}^{\left( 0\right) }\left( x\right) &\propto &\frac{1}{%
U\left( x\right) \sqrt{\sinh \left[ 2\alpha \mu \left( x\right) \right] }}%
\tanh ^{-\frac{2i}{\alpha }}\left[ \alpha \mu \left( x\right) \right] . 
\TCItag{42.d}
\end{eqnarray}

The above models are displayed in their usual forms and give quite
well-known exact solvable non-Hermitian effective potentials as well as
their accompanying ground-state wavefunctions. The first one represents a
generalized $\eta -$weak-pseudo-Hermitian $3D-$harmonic oscillator. The
second model corresponds to the non$-\mathcal{PT-}$symmetric Morse potential
and is already obtained by [22,23], where the $\gamma =b_{R}$ constraint is
considered therein, using $\mathfrak{sl}\left( 2,%
\mathbb{C}
\right) $ potential algebra as a complex Lie algebra by a simple
complexification of the coordinates in a group theoretical point of view and
also in [24], labelled\textbf{\ }LIII according to L\'{e}vai [25], once a
substitution $b\rightarrow ib$ is made therein. The remainder models belong
to so called PI class [25] which contains five individual potentials. The
third model represents a generalized $\eta -$weak-pseudo-Hermitian $\mathcal{%
PT}-$symmetric Scarf II Potential, labelled PI$_{\text{1}}$, which is
established in [22,23,24] with the same constraints quoted above. Finally,
the two last models represent, respectively, a generalized $\eta -$%
weak-pseudo-Hermitian generalized P\"{o}schl-Teller (PI$_{\text{2}}$) and a
generalized $\eta -$weak-pseudo-Hermitian P\"{o}schl-Teller (PI$_{\text{5}}$%
) potentials and are already established, respectively, in [22,23,24] and
[24].

\section{Conclusion}

A well-known class of non-Hermitian Hamiltonians endowed with
position-dependent mass are generated as a by-product of a generalized $\eta
-$weak-pseudo-Hermiticity thanks to a shift on the momentum $p$ of the type $%
p\rightarrow p-\frac{A\left( x\right) }{U\left( x\right) }$, and which
allows to avoid the Hermitian invertible linear operator $\eta $ for the
benefit of $\widetilde{\eta }$. We show that, being different from the
realization of Ref.[13], there is no inconsistency to generate a well-known
class of non-Hermitian Hamiltonians if the last shift is used, leading then
to consider that $\widetilde{\mathcal{D}}$ may be looked upon as a
gauge-transformed version of $\mathcal{D}$ and depending essentially on the
function $A\left( x\right) $, i.e. $\delta \mathcal{D\equiv }\widetilde{%
\mathcal{D}}\mathcal{-D}=-iA\left( x\right) $. As a consequence of this, the
wavefunction $\mathcal{\xi }\left( x\right) $ is also subjected to a gauge
transformation in the manner $\psi \left( x\right) \rightarrow \mathcal{\xi }%
\left( x\right) =\Lambda \left( x\right) \psi \left( x\right) $, with $%
\Lambda \left( x\right) =\exp \left[ i\dint\nolimits^{x}dy\frac{A\left(
y\right) }{U\left( y\right) }\right] $ and where $\psi \left( x\right) $ is
the ground-state wavefunction when the $A\left( x\right) =0$ constraint
holds.

\end{document}